\documentclass[a4paper,11pt]{article}
\pdfoutput=1

\usepackage{jcappub}
\usepackage{subcaption}
\usepackage{units}
\usepackage{siunitx}
\usepackage{mathrsfs}
\usepackage{mathtools}
\usepackage{bm}
\usepackage{comment}
\usepackage[normalem]{ulem}
\usepackage[T1]{fontenc}
\usepackage{lineno}

\title{\boldmath Constraining branon dark matter from observations of the Segue 1 dwarf spheroidal galaxy with the MAGIC telescopes}

\author[a]{T. Miener}
\author[a]{D. Nieto}
\author[b]{V. Gammaldi}
\author[c]{D. Kerszberg}
\author[c]{J. Rico}

\affiliation[a]{Instituto de F\'{i}sica de Part\'{i}culas y del Cosmos and Department of EMFTEL, UCM, E-28040 Madrid, Spain}
\affiliation[b]{Departamento de F\'{i}sica Te\'{o}rica, Universidad Aut\'{o}noma de Madrid, Madrid, Spain $\&$  Instituto de F\'{i}sica Te\'{o}rica, UAM/CSIC, E-28049 Madrid, Spain}
\affiliation[c]{Institut de F\'isica d'Altes Energies (IFAE), The Barcelona Institute of Science and Technology (BIST), E-08193 Bellaterra (Barcelona), Spain}

\emailAdd{tmiener@ucm.es}
\emailAdd{d.nieto@ucm.es}
\emailAdd{viviana.gammaldi@uam.es}
\emailAdd{dkerszberg@ifae.es}
\emailAdd{jrico@ifae.es}

\abstract{
We present the first search for signatures of brane-world extra-dimensional dark matter (DM) in the very-high-energy gamma-ray band by scrutinizing observations of the dwarf spheroidal galaxy Segue 1 with the Major Atmospheric Gamma Imaging Cherenkov (MAGIC) telescope system. Branons are new degrees of freedom that appear within flexible brane-world models: they are weakly interacting massive particles and natural DM candidates. The ground-based gamma-ray telescopes MAGIC could indirectly detect branon DM in the multi-TeV mass range by observing secondary products of DM annihilation into Standard Model particles. In the absence of a signal, we place constraints on the branon DM parameter space by using a binned likelihood analysis of almost 160-hours deep exposure on the Segue 1 dwarf spheroidal galaxy by the MAGIC telescopes. Our most stringent limit to the thermally-averaged annihilation cross-section (at $95\%$ confidence level) corresponds to $ \langle \sigma v \rangle \simeq \unit[1.4 \times 10^{-23}]{cm^{3}s^{-1}} $ at a branon mass of $ \sim \unit[0.7]{TeV}$.}

\begin{document}
\maketitle
\flushbottom

\section{Introduction}
\label{sec:intro}

Astrophysical and cosmological evidences suggest that non-baryonic cold  DM constitutes $84\%$ of the matter density of the Universe~\citep{Aghanim:2018eyx}. Nonetheless, the nature of DM is still an open question for modern physics. In the particle DM paradigm, this elusive kind of matter can not be made of any of the Standard Model (SM) particles. Many efforts have been made in order to determine the nature of the DM, and many candidates have been proposed so far emerging from diverse theories~\citep{Bertone:2005xv}. 

Among others, brane-world theory has been put forward as a prospective framework for DM candidates~\citep{2003PhRvL..90x1301C}. In this theory, the characteristics of the suggested massive brane fluctuations (branons) match the ones of weakly interacting massive particles (WIMPs), which are a well-motivated and widely considered class of cold DM candidates~\citep{2012PhRvD..86b3506S}. WIMPs presenting interaction cross-sections typical of the weak scale would naturally provide the required DM relic density (the so-called {\it WIMP miracle}, see e.g.~\citep{2014IJMPS..3060256S}).

Interacting branons may annihilate into SM particles, consequently rendering them susceptible to be detected. The products of branon annihilation may in turn produce photons e.g. via quark hadronization or final state radiation from charged particles, opening the door to detecting branon annihilation signatures by observing astrophysical regions presenting large DM densities. Given the TeV-mass scale ($ \sim \unit[10]{GeV} $ up to $ \sim \unit[100]{TeV} $) of WIMPs, the MAGIC telescopes, sensitive to very-high energy (VHE, $ \gtrsim \unit[50]{GeV} $) gamma-rays, are an excellent tool to probe branon DM in the multi-TeV mass range.

One of the most promising targets for indirect DM searches are dwarf spheroidal galaxies (dSphs). The dSph satellites, orbiting the Milky Way, are usually less than a few hundred kpc away and have high mass-to-light ratios. In general, these nearby galaxies are less extended, have better determined DM content, and contain less astrophysical background than other DM sources, like the Galactic Center (GC) and galaxy clusters (see, e.g,~\citep{2002PhRvL..88s1301M,2007ApJ...654..897B}). In this work, we are focusing on the dSph galaxy Segue 1. 

This article is structured as follows: Section \ref{sec:BranonDM} succinctly introduces the main features of the brane-world theory and the expected photon flux from branon DM annihilation; the observational campaign on Segue 1 by the MAGIC telescopes is presented in Section \ref{sec:ObservationsAnalysis} as well as the adopted analysis methodology. In Section \ref{sec:Results}, we present the first upper limits to the annihilation cross section of branon DM particles using very-high-energy gamma-ray observations. We then discuss our results and finally present our conclusions in Section \ref{sec:DisCon}.

\section{Branon dark matter}
\label{sec:BranonDM}

\subsection{Brane-world theory}

The framework of extra-dimensional models~\citep{1998PhLB..429..263A,1999PhRvD..59h6004A}, theorizes that the SM fields exist on a tridimensional brane embedded into a higher dimensional spacetime - with $D$ dimensions, where $D=4+N$ and $N$ is the number of extra dimensions - where gravity propagates. These models were proposed as a potential solution to the hierarchy problem, but also provide us with natural DM particle candidates. In the particular context of the so-called \textit{brane-world scenario}, and for low brane tension as compared to the fundamental scale of gravity, branons are massive brane fluctuations in the direction of the N-extra-dimensions whose relic abundance can account for the cosmological DM~\citep{2003PhRvL..90x1301C,2003PhRvD..68j3505C}. The lowest-order effective Lagrangian for branon dark matter (BDM) reads~\citep{2003PhRvD..67g5010A, 2012PhRvD..85d3505C}

\begin{equation}
\label{eq:lagrangian}
    \mathscr{L}_{\text{BDM}} = \frac{1}{2} g^{\mu\nu} \partial_{\mu} \pi^{\alpha} \partial_{\nu} \pi^{\alpha} - \frac{1}{2} m_{\chi}^{2}\pi^{\alpha}\pi^{\alpha} + \frac{1}{8f^{4}} \left( 4 \partial_{\mu} \pi^{\alpha} \partial_{\nu} \pi^{\alpha} - m_{\chi}^{2}\pi^{\alpha}\pi^{\alpha} g_{\mu\nu} \right) T_{\text{SM}}^{\mu\nu},
\end{equation}
\noindent
where $\pi$ denotes the branon field and ${\alpha}$ runs over the number of extra dimensions $N$, $f$ and $m_{\chi}$ are the tension of the brane and the mass of the branon respectively, and $T^{\mu\nu}_{\text{SM}}$ is the energy-momentum tensor of the SM fields. As can be seen in Eq.~\ref{eq:lagrangian}, the coupling of the branons to the SM particles is suppressed by the fourth power of the tension of the brane, rendering them as weakly interacting particles. In the simplest case of this effective field theory, there is only one extra-dimension, i.e. $\alpha=1$, and thus a single branon particle. 

Branons may self-annihilate into SM particles. For non relativistic branons, the leading term in the thermally averaged cross section of annihilation into, respectively: Dirac fermions~$ \psi $ with mass $ m_{\psi} $, massive gauge fields ($ W $ or $ Z $) with mass $ m_{W,Z} $ and (complex) scalar field~$ \Phi $ with mass $ m_{\Phi} $ can be expressed as~\citep{2012PhRvD..85d3505C}

\begin{align}
\label{eq:Cross_sections_psi}
\langle \sigma_{\psi} v \rangle &= \frac{m_{\chi}^{2}m_{\psi}^{2}}{16\pi^{2}f^{8}}\left( m_{\chi}^{2}-m_{\psi}^{2} \right) \sqrt{1-\frac{m_{\psi}^{2}}{m_{\chi}^{2}}}, \\
\label{eq:Cross_sections_wz}
\langle \sigma_{W,Z} v \rangle &= \frac{m_{\chi}^{2}}{64\pi^{2}f^{8}}\left( 4m_{\chi}^{4} - 4m_{\chi}^{2}m_{W,Z}^{2} + 3 m_{W,Z}^{4} \right) \sqrt{1-\frac{m_{W,Z}^{2}}{m_{\chi}^{2}}}, \\
\label{eq:Cross_sections_phi}
\langle \sigma_{\Phi} v \rangle &= \frac{m_{\chi}^{2}}{32\pi^{2}f^{8}}\left( 2m_{\chi}^{2}  + m_{\Phi}^{2} \right)^{2} \sqrt{1-\frac{m_{\Phi}^{2}}{m_{\chi}^{2}}}.
\end{align}
\noindent
By considering the annihilation in quark channels, a factor 3 was required in \ref{eq:Cross_sections_psi}, in order to take into account the three different quark colors (with respect to the annihilation in leptonic channels); since the massive gauge field $ W $ is complex, an additional factor of 2 was added to \ref{eq:Cross_sections_wz} for the $ W^{+}W^{-}$ annihilation channel; finally, a factor 1/2 was included in Eq. \ref{eq:Cross_sections_phi}, to consider that the Higgs boson is a real (non-complex) scalar field. For a massless gauge field~$ \gamma $, the leading order of the cross section is essentially zero,  since this is a \emph{d-wave} annihilation process and is thus highly suppressed. 

The branching ratios into each possible annihilation channel can then be expressed as 

\begin{equation}
    \label{eq:Branching_ratios}
    \text{BR}_{i}\left( m_{\chi} \right)  = \frac{\langle \sigma_{i} v \rangle}{\langle \sigma v \rangle}\text{, with }\langle \sigma v \rangle = \sum \mathop{}_{\mkern-5mu j} \langle \sigma_{j} v \rangle.
\end{equation}
\noindent
It can be seen from Eqs.~\ref{eq:Cross_sections_psi} to~\ref{eq:Cross_sections_phi} that the mass of the branon is the only variable determining the values of the branching ratios, $\text{BR}_{i}$, whereas the total thermally-averaged annihilation cross section, $\langle \sigma v \rangle$, depends on both the tension of the brane and the mass branon.  

\subsection{Expected branon dark matter flux}
\label{sec:ExpBranonDMFlux}

In our derivations below we assume a single extra-dimension and therefore a single type of branon. The expected differential photon flux produced by branon DM annihilation in a given region of the sky, $\Delta\Omega$, and observed at Earth can be expressed as
\begin{equation}
    \label{eq:Branon_Flux}
    \frac{\text{d}\Phi_{\text{BDM}}}{\text{d}E}\left( \Delta\Omega,\langle\sigma v\rangle \right) = J \cdot \frac{1}{4\pi} \frac{\langle\sigma v\rangle}{2m^{2}_{\chi}}  \frac{\text{d}N_{\text{BDM}}}{\text{d}E},
\end{equation}
\noindent
where $ \langle\sigma v\rangle $ and $ m_{\chi} $ are, respectively, the thermally-averaged annihilation cross section and the mass of the branon DM particle as previously introduced, and 
\begin{equation}
    \label{eq:Branon_dNdE}
    \frac{\text{d}N_{\text{BDM}}}{\text{d}E} = \sum_{i=1}^{n} \text{BR}_{i} \frac{\text{d}N_{i}}{\text{d}E}
\end{equation}
\noindent 
is the differential photon yield per annihilation, which is a weighted sum over all the $ n $ possible SM annihilation channels whose products can produce photons. All the information regarding the spectral shape of the gamma-ray flux produced by branon DM annihilation (see right panel of Fig.~\ref{fig:Branching_ratios_and_spectra}) is contained in the $ \text{d}N_{\text{BDM}}/\text{d}E $ term.

The \textit{astrophysical} factor (\textit{J}-factor) depends on both the distance and the DM distribution at the source region. It is given by
\begin{equation}
    J = \int_{\Delta\Omega} d\Omega' \int_{\text{l.o.s.}} dl \, \rho_{\text{DM}}^{2} (l,\Omega'),
\end{equation}
\noindent
where $ \text{l.o.s.} $ stands for line-of-sight and $ \rho_{\text{DM}} $ is the DM density.

\section{Observations and analysis method}
\label{sec:ObservationsAnalysis}

\subsection{Segue 1 observation by the MAGIC telescopes}
\label{subsec:Segue1ObservationsMAGIC}
The ultra-faint dSph Segue 1, of absolute magnitude $ M_{V} = -1.5_{-0.8}^{+0.6} $, was discovered in the Sloan Digital Sky Survey (SDSS) imaging data in 2006~\citep{2007ApJ...654..897B} and is with an estimated $ \sim 3400 \; M_{\odot}/L_{\odot} $ mass-to-light ratio one of the most DM-dominated object known so far~\citep{2011ApJ...733...46S}. Besides, Segue 1 is positioned in the Northern Hemisphere and outside of the Galactic plane ($ \text{RA} = \unit[10.12]{h} $, $ \text{DEC} = 16.08^{\circ}$) only $ \unit[23\pm2]{kpc}$ away from us, which leads to an excellent target for indirect DM searches in the VHE gamma-ray bands with the MAGIC telescopes~\citep{2011JCAP...06..035A,2012PhRvD..85f2001A,2014JCAP...02..008A,2016JCAP...02..039M}.

The MAGIC telescopes consist of a system of two 17 m diameter telescopes operating in stereoscopic mode at the Roque de los Muchachos Observatory ($ 28.8^{\circ}$ N, $ 17.9^{\circ}$ W; 2200~m~a.s.l.) on the Canary Island of La Palma, Spain. The fast imaging cameras with a field of view of $ 3.5^{\circ} $, installed in the two telescopes, detect the Cherenkov light produced by the atmospheric showers initiated by cosmic particles entering the Earth atmosphere. The system is able to identify and reconstruct cosmic gamma-ray events in the VHE domain~\citep{2016APh....72...76A}.

In our analysis, we use the stereoscopic observation of the dSph galaxy Segue 1 with MAGIC\footnote{The observation performed by the MAGIC-I telescope in single telescope mode~\citep{2011JCAP...06..035A} are not included in our analysis.}, which were already described and analyzed in~\citep{2014JCAP...02..008A,2016JCAP...02..039M}. This observational campaign was carried out between 2011 and 2013 and is with 157.9 h the deepest observation of any dSph by a Cherenkov telescope to date.

\subsection{Likelihood analysis}
\label{subsec:LklAnalysis}

The data reduction of the Segue 1 observation have been kindly provided by the MAGIC Collaboration. It was performed with the standard MAGIC analysis software MARS~\citep{2013ICRC...33.2937Z} and is exactly the same as for~\citep{2014JCAP...02..008A,2016JCAP...02..039M}. In this project, we re-analyse those high-level data in the context of brane-world extra-dimensional theories using \texttt{gLike}~\citep{javier_rico_2021_4601451} and \texttt{LklCom}~\citep{tjark_miener_2021_4597500}. In particular, we are using a DM-oriented approach for our likelihood analysis that takes the expected signal spectral shape of the specific DM model into account. Aleksi\'{c}, Rico and Martinez have shown in~\citep{2012JCAP...10..032A} that this approach significantly improves the sensitivity to gamma-ray signals of DM origin with respect to a Poisson likelihood approach. Different to~\citep{2014JCAP...02..008A}, we are using a binned likelihood function and include the systematic uncertainty on the residual background contamination in our analysis as described in the following. The same binned version of the likelihood analysis is being used by the MAGIC Collaboration to produce limits to the annihilation of generic WIMPs using all dSph observations by MAGIC~\citep{2021arXiv211115009M} and the current generation of gamma-ray instruments \textit{Fermi}-LAT, HESS, MAGIC, VERITAS and HAWC~\citep{Oakes:2019, Armand:2021}.

We model the gamma-ray emission in the source region with the branon DM model and then compare the expected spectral distribution to the measured one. Since the spectral shape is known for the model (see Section \ref{sec:BranonDM}), the intensity of the gamma-ray signal $ \langle \sigma v \rangle $ is the only free parameter. The corresponding binned ($ N_{\text{bins}} = 30 $) likelihood function of the dataset $ \bm{\mathcal{D}} $ with nuisance parameters $ \bm{\mu} $ can be written as:
\begin{equation}
    \label{eq:Binned_lkl}
    \begin{split}
    \mathcal{L}_{\text{bin}} \left( \langle \sigma v \rangle ; J, \bm{\mu} \mid \bm{\mathcal{D}} \right) &= \mathcal{L}_{\text{bin}}( \langle \sigma v \rangle; \{ b_{i} \}_{i=1,\ldots,N_{\text{bins}}}, J, \tau \mid \{ N_{\text{ON},i}, N_{\text{OFF},i} \}_{i=1,\ldots,N_{\text{bins}}}) \\
    &= \prod_{i=1}^{N_{\text{bins}}} \Big[ \mathcal{P} (s_{i}(\langle \sigma v \rangle, J) + b_{i} \mid N_{\text{ON},i}) \cdot \mathcal{P} (\tau b_{i} \mid N_{\text{OFF},i}) \Big] \times \mathcal{T} \left( \tau \mid \tau_{\text{o}}, \sigma_{\tau} \right)
    \end{split}
\end{equation}
\noindent
where $ \mathcal{P} (x | N) $ is the Poisson distribution of mean $x $ and measured value $ N $ and $ s_{i}(\langle \sigma v \rangle, J) $ (see Eq. \ref{expected_numbers_of_signal_events}) and $ b_{i} $ are the expected numbers of signal and background events in the $ i $-th energy bin. The total number of observed events in a given energy bin $ i $ in the signal (ON) and background (OFF) regions are $ N_{\text{ON},i} $, $ N_{\text{OFF},i} $, respectively. The normalization between background and signal regions is denoted with $ \tau $. Besides the expected number of background events $ b_{i} $, $ \tau $, described by the likelihood function $ \mathcal{T} $, is also a nuisance parameter in the analysis. $ \mathcal{T} \left( \tau \mid \tau_{\text{o}}, \sigma_{\tau} \right) $ is a Gaussian function with mean $ \tau_{\text{o}} $ and variance $ \sigma_{\tau}^{2} $, which include statistical and systematics uncertainties. We considered a systematic uncertainty of $ \sigma_{\tau_{\mathrm{syst}}} = 1.5\%  \cdot \tau $ on the estimate of the residual background based on the dedicated performance study of the MAGIC telescopes~\citep{2016APh....72...76A}.

The expected number of signal events in the $ i $-th energy bin is
\begin{equation}
    \label{expected_numbers_of_signal_events}
    s_{i}(\langle \sigma v \rangle, J) = T_{\text{obs}} \int_{E_{\text{min},i}}^{E_{\text{max},i}} \text{d}E' \int_{0}^{\infty} \frac{\text{d}\Phi_{\text{BDM}}(\langle \sigma v \rangle, J)}{\text{d}E} R_{\text{ON}} \left( E, E'\right) \text{d}E,
\end{equation}
\noindent
where $ T_{\text{obs}} $ is the total observation time, $ E $ and $ E' $ are respectively the true and reconstructed energy, $ E_{\text{min},i} $ and $ E_{\text{max},i} $ are the lower and upper limits of the $ i $-th energy bin, $ \text{d}\Phi_{\text{BDM}}/\text{d}E $ is the expected branon DM flux (Eq. \ref{eq:Branon_Flux}) in the signal region, and $ R_{\text{ON}} \left( E, E'\right) $ is the telescope response function for the signal region, which can be described by the effective collection area ($A_{\text{eff}}$) and by the PDF for the energy estimator.

\begin{figure}[ht]
    \centering
    \begin{subfigure}{.49\textwidth}
        \includegraphics[width=\textwidth]{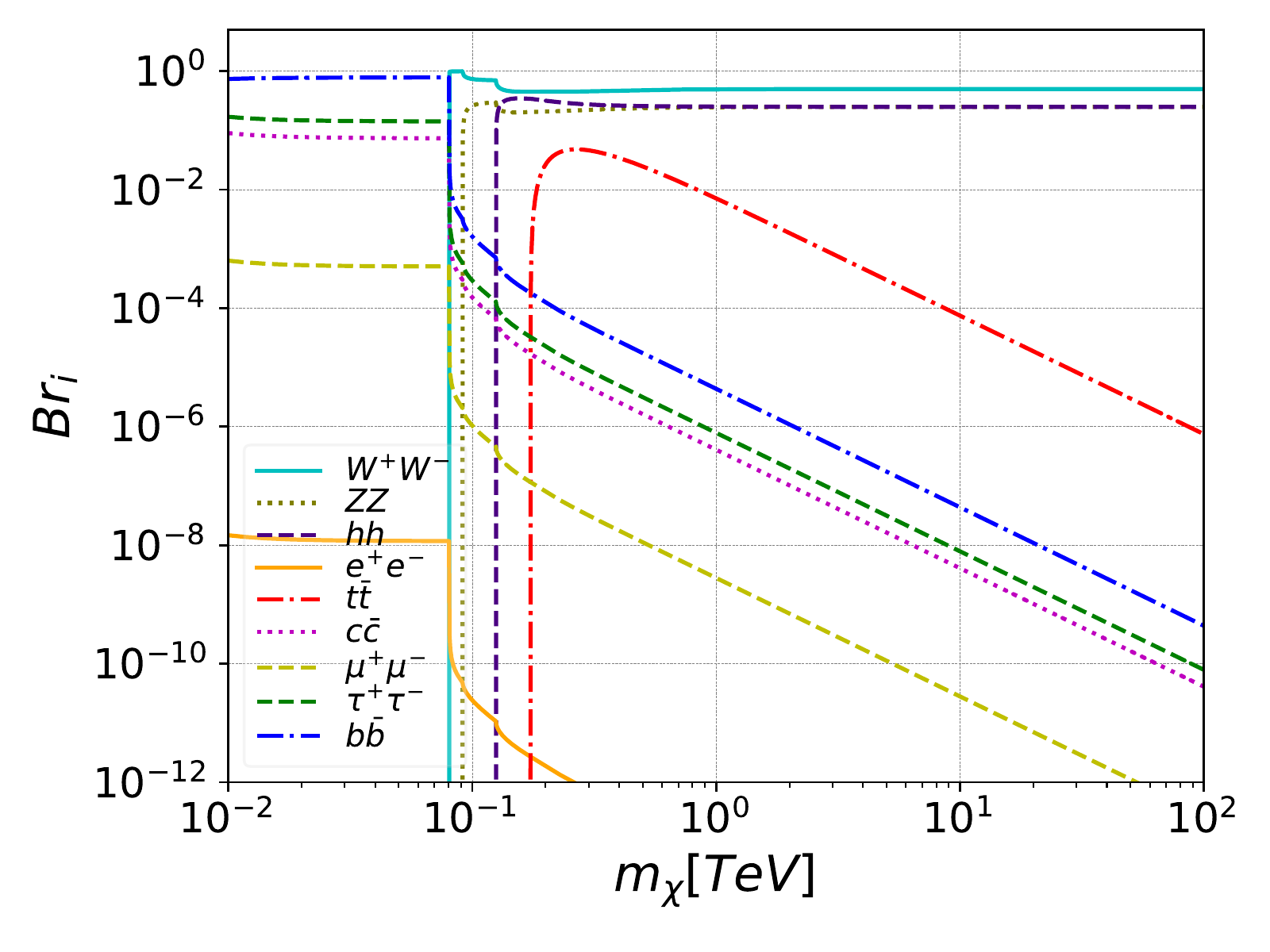}
    \end{subfigure}
    \begin{subfigure}{.49\textwidth}
        \includegraphics[width=\textwidth]{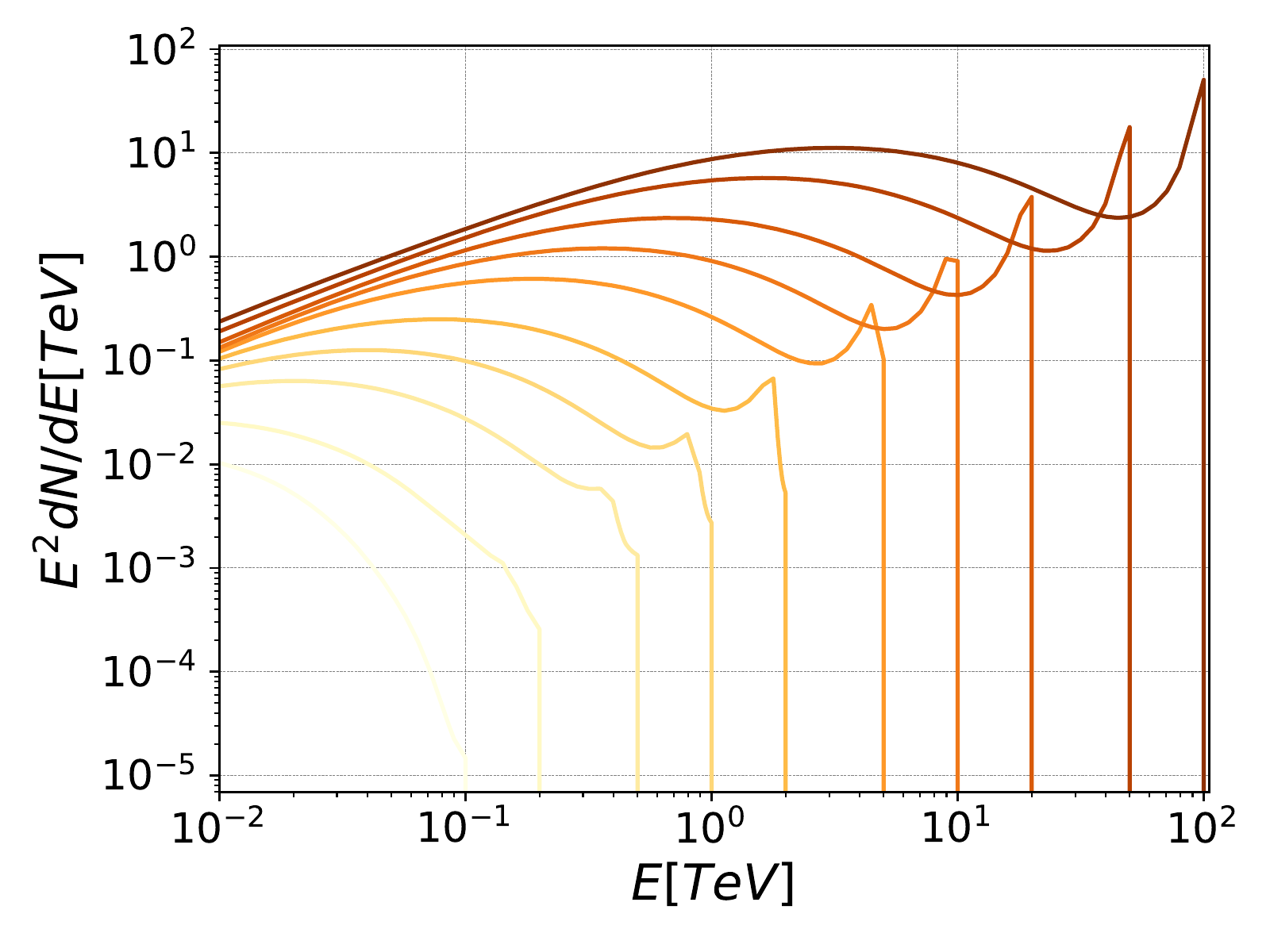}
    \end{subfigure}
\caption{{\em Left:} The branon branching ratios (Eq. \ref{eq:Branching_ratios}) as a function of the only free parameter $ m_{\chi} $ in the expected mass range ($ \unit[10]{GeV} $ up to $ \unit[100]{TeV} $) of WIMPs. {\em Right:} The differential photon yield per branon annihilation $ \text{d}N_{\text{BDM}}/\text{d}E $ (Eq. \ref{eq:Branon_dNdE}) for a set of branon DM masses (from light to dark: 0.1, 0.2,
0.5, 1, 2, 5, 10, 20, 50 and $ \unit[100]{TeV} $).}
\label{fig:Branching_ratios_and_spectra}
\end{figure}

In our analysis, we are calculating the branon branching ratios $ \text{BR}_{i} $ (Eq. \ref{eq:Branching_ratios}) including annihilation into the SM pairs $ W^{+}W^{-} $, $ ZZ $, $ hh $, $ e^{+}e^{-} $, $ t\bar{t} $, $ c\bar{c} $, $ \mu^{+}\mu^{-} $, $ \tau^{+}\tau^{-} $ and $ b\bar{b} $ (see left panel of Fig. \ref{fig:Branching_ratios_and_spectra}). The $ b\bar{b} $ channel dominates for lower masses, while for masses above $ \unit[80]{GeV} $ the $ W^{+}W^{-} $ channel has the largest impact~\citep{2012PhRvD..85d3505C}. Hence, given our energy sensitivity, the $ W^{+}W^{-} $, $ ZZ $ and $ hh $ channels are the most significant contributors in our analysis. The  differential gamma-ray yields per annihilation $ \text{d}N_{i}/\text{d}E $ are taken from the PPPC 4 DM ID distribution~\citep{2011JCAP...03..051C}.

The Segue 1 observational campaign $ \bm{\mathcal{D}}_{\mathrm{Segue1}} $ results in $ N = 4 $ distinct datasets\footnote{Due to major hardware upgrade~\citep{2016APh....72...61A}, the data set is divided into four different observation periods and each period is treated with an individual set of instrument response functions (IRFs). Besides that the data were taken in wobble mode~\citep{1994APh.....2..137F} with two pointing (wobble) positions, which leads to eight samples in total.}. The joint likelihood function
\begin{equation}
    \mathcal{L}\left( \langle \sigma v \rangle ; J, \bm{\nu} \mid \bm{\mathcal{D}}_{\text{Segue1}} \right) = \prod_{k=1}^{N} \Big[ \mathcal{L}_{\text{bin},k} \left( \langle \sigma v \rangle ; J, \bm{\nu_{k}} \mid \bm{\mathcal{D}_{k}} \right) \Big] \times \mathcal{J} \left( J \mid J_{\text{o}}, \sigma_{\log_{10}J} \right)
\end{equation}
\noindent
is the product of the likelihood function of each dataset. We treat the \textit{J}-factor as a nuisance parameter and include the likelihood $ \mathcal{J} $ for the \textit{J}-factor following~\citep{2015PhRvL.115w1301A}. $ \bm{\nu_{k}} $ represents the set of nuisance parameters different from the \textit{J}-factor affecting the analysis of the $ k $-th dataset.

In our analysis, we are using the \textit{J}-factor and its statistical uncertainty for Segue 1 from~\citep{2015ApJ...801...74G}, where the DM density distribution is modeled assuming a Navarro-Frenk-White (NFW) DM density profile~\citep{1996ApJ...462..563N}. Thus, we consider the value of $ \mathrm{log}_{10} \left( J \left( \theta \right) [\unit{GeV^{2} cm^{\num{-5}}}] \right) = 19.02_{\num{-0.35}}^{\num{+0.32}} $ integrated up to the angular distance $ \theta = 0.125^{\circ} $ of the Segue 1 DM halo according to the signal region.

\section{Results}
\label{sec:Results}

\begin{figure}[h]
    \centering
    \includegraphics[scale=0.7]{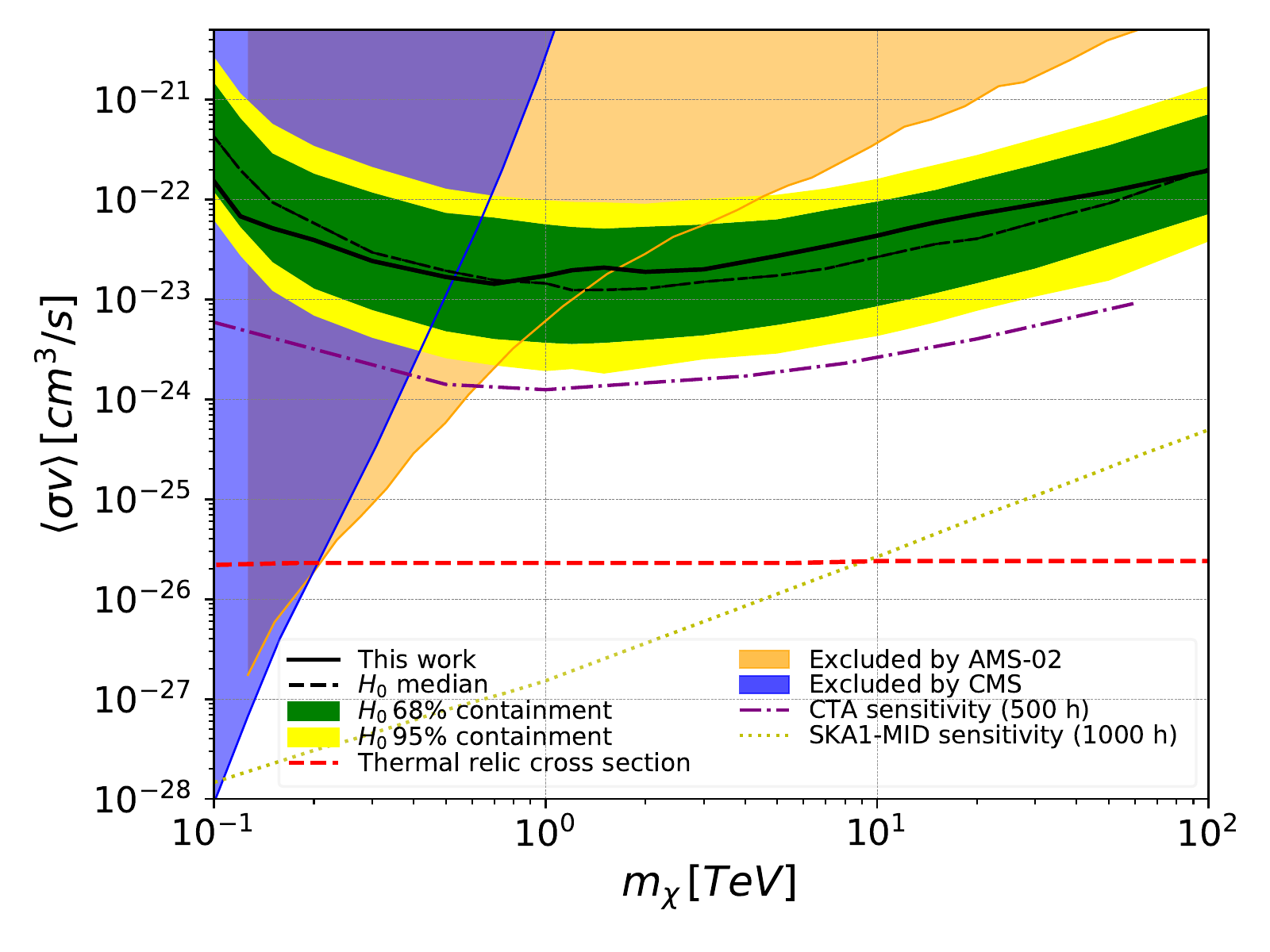}
    \caption{The $ 95 \% $ CL upper limits on $ \langle \sigma v \rangle $ for branon DM annihilation. The solid black line shows our branon limits, while the dotted black line, green and yellow bands show the median and the two-sided $ 68 \% $ and $ 95 \% $ containment bands, respectively. The thermal relic cross-section from~\citep{2012PhRvD..86b3506S} is indicated by the red-dashed line. The tightest constraints to branons model by colliders are obtained from CMS data and represented by the blue exclusion region~\citep{2014arXiv1410.8812C}. The analysis of AMS-02 $ e^{+}e^{-} $ data excludes the orange region~\citep{2017arXiv170909819C}. Both exclusion regions were translated to the $ \langle \sigma v \rangle $ parameter space from~\citep{2020PDU....2700448C}. The purple dashed-dotted line represents the estimated branon sensitivity for 500 h observation on the dSph Draco with the future CTA~\citep{2020JCAP...10..041A}. The estimated sensitivity for 1000 h observation on the classical dSph Draco with the planned SKA, assuming the $ W^{+}W^{-} $ annihilation mode, are represented by the yellow dotted line~\citep{2020PDU....2700448C}.}
    \label{fig:BranonLimits}
\end{figure}

We present the first observational $ 95 \% $ confidence level (CL) upper limits on the thermally-averaged cross-section $ \langle \sigma v \rangle $ (see Fig. \ref{fig:BranonLimits}), in the context of brane-world extra-dimensional theories, obtained with 157.9 hours of good quality stereoscopic data from the Segue 1 observation with the MAGIC telescopes. These limits were computed by following the prescription from~\citep{2015PhRvL.115w1301A, 2016JCAP...02..039M}, with $ \langle \sigma v \rangle $ restricted to the physical region ($ \langle \sigma v \rangle \geqslant 0 $). Different from previous works~\citep{2016JCAP...02..039M, 2015PhRvL.115w1301A, 2018JCAP...03..009A, 2011JCAP...06..035A, 2014JCAP...02..008A}, we performed a model dependent search for branon DM particles of masses between $ \unit[100]{GeV} $ and $ \unit[100]{TeV} $. The final results were computed assuming no additional boosts from the presence of substructures~\citep{2007PhRvD..75h3526S} or quantum effects~\citep{2004PhRvL..92c1303H}.

We used a binned likelihood analysis, including systematic uncertainties in the residual background intensity and statistical uncertainties in the \textit{J}-factor, to set these first constraints on the branon DM model from gamma-ray observations. The two-sided $ 68 \% $ and $ 95 \% $ containment bands as well as the median were estimated from the distribution of the upper limits obtained when performing the same analysis of 300 fast simulations of the source and background regions assuming no DM signal ($ \langle \sigma v \rangle = 0 $).

\begin{figure}[h]
    \centering
    \includegraphics[scale=0.7]{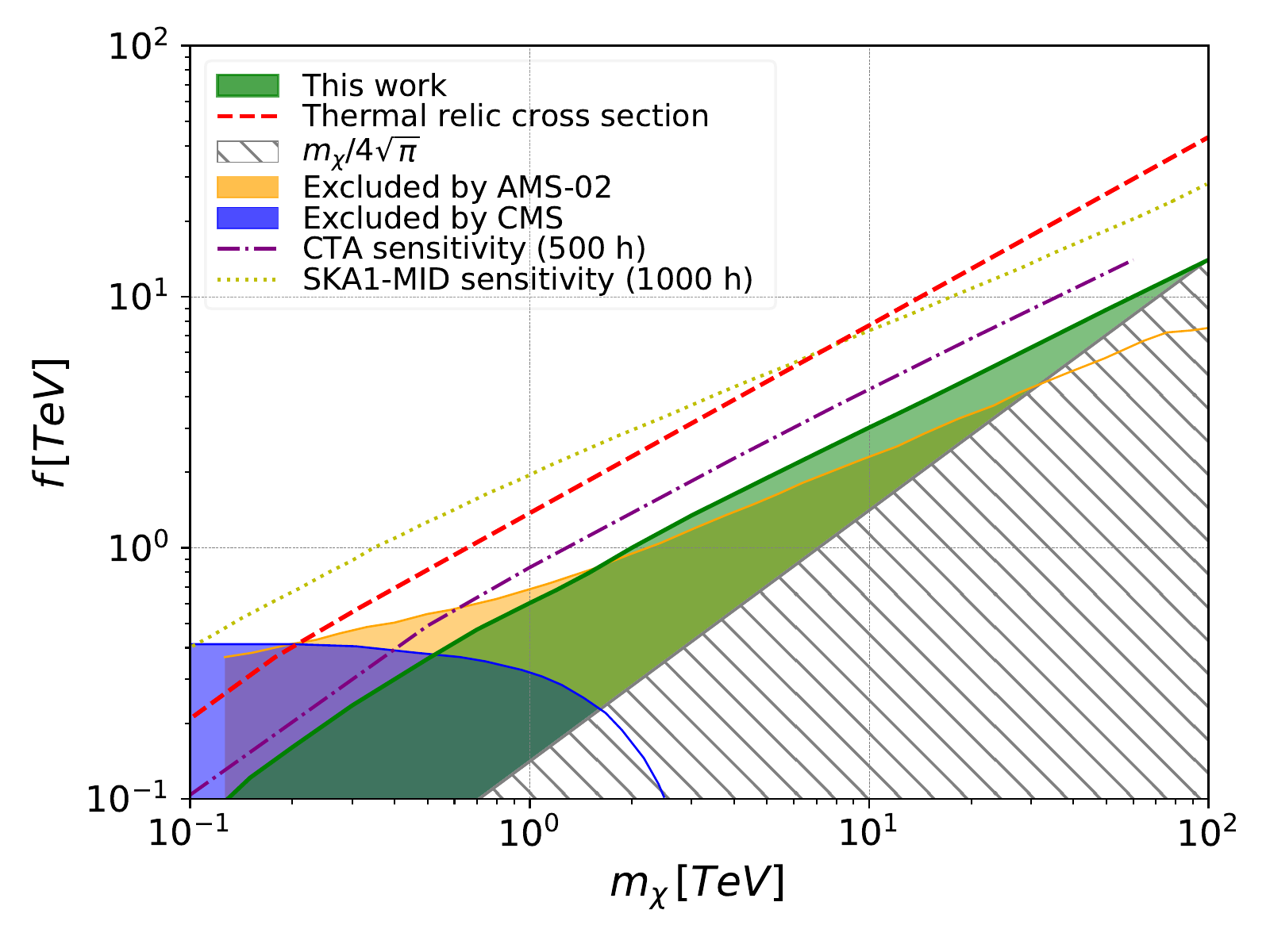}
    \caption{The $ 95 \% $ CL upper limits on the brane tension $ f $ for branon DM annihilation. Our branon limits are depicted by the green exclusion region. The thermal relic cross-section from~\citep{2012PhRvD..86b3506S} is indicated by the red-dashed line. The tightest constraint to branons model by colliders are obtained from CMS data and represented by the blue exclusion region~\citep{2014arXiv1410.8812C}. The analysis of AMS-02 $ e^{+}e^{-} $ data excludes the orange region~\citep{2017arXiv170909819C}. Both exclusion regions were taken from~\citep{2020PDU....2700448C}. The model validity limit in the $f(m_\chi)$ parameter space is depicted by the grey dashed region. The purple dash-dotted line represents the estimated branon sensitivity for 500 h observation on the dSph Draco with the future CTA~\citep{2020JCAP...10..041A}. The estimated sensitivity to branons for 1000 h observation on the dSph Draco with the planned SKA are represented by the yellow dotted line~\citep{2020PDU....2700448C}.}
    \label{fig:Branetension_limits}
\end{figure}

Our constraints are located within the  $ 68 \% $ containment band, which is consistent with the no-detection scenario. As already reported in~\citep{2011JCAP...06..035A,2012PhRvD..85f2001A,2014JCAP...02..008A,2016JCAP...02..039M}, no significant gamma-ray excess has been found in the Segue 1 data. Our strongest limit is $ \langle \sigma v \rangle \simeq \unit[1.4 \times 10^{-23}]{cm^{3}s^{-1}} $ for a $ \sim \unit[0.7]{TeV} $ mass branon DM particle. Differently from model independent DM searches, we are able to set constraints to a specific parameter space of the branon DM model, i.e. the tension of the brane $ f $ versus the DM mass. In fact, since the total annihilation cross section $ \langle \sigma v \rangle = \sum_{j}\langle \sigma_{j} v \rangle $ only depends on $ m_{\chi} $ and $ f $, we can translate our $ \langle \sigma v \rangle $ limits to constraints on $ f $. This allows us to exclude a significant portion of the brane tension versus branon mass parameter space, $f(m_\chi)$, ranging from 0.1 to 100 TeV in branon mass, as shown in figure~\ref{fig:Branetension_limits}. We note that MAGIC enlarges the region of the parameter space that has already been excluded by AMS-02~\citep{2017arXiv170909819C} and CMS~\citep{2014arXiv1410.8812C}, especially for branon masses above 1 TeV. In future, the Cherenkov Telescope Array (CTA) and the Square Kilometer Array (SKA) will probe a larger fraction of the exclusion region, providing valuable complementary information in both gamma-ray and radio observations, respectively. 

\section{Discussion and conclusions}
\label{sec:DisCon}

We have reported the indirect search for branon DM in the dSph galaxy Segue 1 using the MAGIC telescopes data. This observational campaign is still with 157.9 hours the deepest survey of any dSph by any imaging atmospheric Cherenkov telescope to date. The data of each observation period have been analyzed by means of the binned likelihood method, taking the spectral shape from branon DM annihilation into account. Subsequently, the likelihood functions of each dataset were combined in a joint analysis, which is treating the normalization between background and signal regions $ \tau $ and the \textit{J}-factor as nuisance parameters in the likelihood.

Above $ \sim \unit[1]{TeV}$, this work is superseding the limits previously obtained from analysis from AMS-02~\citep{2017arXiv170909819C} and CMS~\citep{2014arXiv1410.8812C} and leading to the most constraining branon DM limits in the multi-TeV mass range. Even more stringent exclusion limits of the branon DM annihilation can be achieved by combining further dSph observations of the MAGIC telescopes~\citep{2021arXiv211115009M}. In the framework of multi-instrument and multi-messenger DM searches~\citep{Oakes:2019, Armand:2021} a global branon DM limit over a wider range of DM masses can be obtained with a joint analysis of observational data from different gamma-ray and neutrino telescopes.

\acknowledgments
The authors thank the MAGIC Collaboration for providing private data.\newline
DN and TM acknowledge support from the former {\em Spanish Ministry of Economy, Industry, and Competitiveness / European Regional Development Fund} grant FPA2015-73913-JIN.\newline
VG's contribution to this work has been supported by Juan de la Cierva-Formaci\'on FJCI-2016-29213 and Juan de la Cierva-Incorporaci\'on IJC2019-040315-I grants, the Spanish Agencia Estatal de Investigaci\'on through the grants FPA2015-65929-P (MINECO/FEDER, UE), PGC2018-095161-B-I00 and IFT Centro de Excelencia Severo Ochoa SEV-2016-0597. VG also acknowledges the support of the Spanish Red Consolider MultiDark FPA2017-90566-REDC. VG thanks J.A.R. Cembranos for useful discussions. \newline
DK is supported by the European Union's Horizon 2020 research and innovation programme under the Marie Sk\l{}odowska-Curie grant agreement No. 754510. DK and JR acknowledge the support from the ERDF under the Spanish Ministerio de Ciencia e Innovaci\'{o}n (MICINN, grant PID2019-107847RB-C41), from the Centro de Excelencia Severo Ochoa SEV-2016-0588, and from the CERCA program of the Generalitat de Catalunya.

\bibliographystyle{JHEP}
\bibliography{BibTex}

\end{document}